\documentstyle[osa,manuscript,psfig]{revtex}

\begin{document}

\titlepage
\title{Spin structure and 
longitudinal polarization of hyperon 
in $e^+e^-$ annihilation at high energies} 
\author {Liu Chun-xiu, and Liang Zuo-tang}
\address {Department of Physics,
Shandong University,Jinan, Shandong 250100,China}

\maketitle

\begin{abstract}     

Longitudinal polarizations of different 
kinds of hyperons produced in 
$e^+e^-$ annihilation at LEP I 
and LEP II energies in different event samples 
are calculated using two different pictures for 
the spin structure of hyperon:
that drawn from polarized deep inelastic 
lepton-nucleon scattering data or 
that using SU(6) symmetric wave functions.
The result shows that measurements of such polarizations 
should provide useful information to the question 
of which picture is more suitable in describing 
the spin effects in the fragmentation processes. 

\end{abstract}     

\newpage

\section{Introduction}

Spin effects in high energy fragmentation processes 
have attracted much attention\cite{att} recently. 
Study of such effects provide useful information 
for the spin structure of hadron and spin dependence 
of high energy reactions.  
There exist now two distinctively different 
pictures for the spin contents of the baryons:
the static quark model picture 
using SU(6) symmetric wave function  
[hereafter referred as SU(6) picture],  
and the picture drawn from the data for polarized deep inelastic 
lepton-nucleon scattering (DIS) \cite{SPIN97}
and SU(3) flavor symmetry in hyperon decay 
[hereafter referred as DIS picture].  
It is natural to ask which picture 
is suitable to describe the relationship 
between the polarization of the fragmenting quark 
and that of the produced hadron. 
It has been pointed out in [\ref{BL98}] that 
measurements of $\Lambda$ longitudinal polarization 
in $e^+e^-$ annihilation at high energies can 
provide useful information to answer this question. 
Calculations based on these 
two different pictures have been made 
in [\ref{GH93}] and [\ref{BL98}].
A comparison of the obtained results 
with the ALEPH data\cite{ALEPH96} obtained at LEP,  
which was available at that time, was made. 
The result shows that the SU(6) picture seems 
to agree better with the data\cite{ALEPH96}. 
This is rather surprising since the energy 
is very high at LEP and the initial 
quarks and anti-quarks produced 
at the annihilation vertices of 
the initial $e^+e^-$ are certainly 
current quarks and current anti-quarks 
rather than the constituent quarks used 
in describing the static properties of hadrons 
using SU(6) symmetric wave functions.
Other models\cite{Kotz98,Ma2000} are also proposed 
which can give a description of the available data\cite{ALEPH96}.
The available data\cite{ALEPH96} 
are certainly still far from accurate and 
abundant enough to make a conclusive judgment 
of these different models. 
It is therefore important and urgent to 
make further checks of this conclusion 
by performing complementary measurements.

In this paper, we formulate the calculation method  
of longitudinal polarization of different hyperons 
in $e^+e^-$ annihilation at high energies proposed 
in [\ref{GH93}] and [\ref{BL98}] in a systematic way 
so that they can be generalized to other hyperons 
and/or other reactions.  
We describe in detail the inputs and/or assumptions 
which have been used in the calculations.
We then present the calculated results 
for different hyperons based on these two pictures. 
We present the results for the whole events 
as well as those for different 
particularly chosen event samples.
We show that corresponding measurements 
can be used as complementary method to 
check the above mentioned conclusion. 
In particular, we discuss also how to distinguish such models 
from the kind of models proposed in [\ref{Kotz98},\ref{Ma2000}].  
In section 2, we outline the calculation method 
and present the results for $\Lambda$ polarization. 
In section 3, we present 
our results for other hyperons. 

\section{The calculation method and the results for $\Lambda$ 
in different event samples} 

We recall that\cite{AR80}, according to the standard model 
of electroweak interactions, quarks and anti-quarks 
produced at the annihilation vertices of 
$e^+e^-$ are longitudinally polarized. 
Such longitudinal quark polarization 
can be transferred to the produced hadron 
which contains the initial quark thus lead 
to longitudinal polarization 
of hadron in the inclusive process
$e^+e^-\to h+X$.  
Measuring the polarizations of the produced hadrons,
we can study the relation between the 
spin of the fragmenting quark and that of 
the produced hadron which contains that quark.
In this aspect, the $J^P={1\over 2}^+$ hyperons are most 
suitable candidates among all the hadrons.
This is because these hyperons decay weakly so that
their polarizations can easily be obtained from
the angular distributions of their decay products.
We now outline the calculation method of 
the longitudinal polarization $P_{H_i}$ of different 
hyperon  $H_i$ in the inclusive process 
$e^+e^-\to  H_i+X$.

\subsection{The calculation method for different hyperons}
\label{subs:method}

Since the longitudinal polarization $P_{H_i}$ 
of the hyperon $H_i$ in the inclusive process  
$e^+e^-\to  H_i+X$ originates from 
the longitudinal polarization $P_f$  
of the initial quark $q^0_f$ 
(where the subscript $f$ denotes its flavor)
produced at the annihilation vertex 
of the initial state $e^+e^-$, 
we should consider the $H_i$'s 
which have the following different origins separately. 

(a) Hyperons which are directly produced 
and contain the 
initial quarks $q_f^0$'s originated from 
the annihilations of the initial $e^+$ and $e^-$;

(b) Hyperons which are decay products of other heavier 
hyperons which were polarized before their decay; 

(c) Hyperons which are directly produced but 
do not contain any initial quark $q_f^0$ from $e^+e^-$ 
annihilation;

(d) Hyperons which are decay products of other heavier hyperons 
which were unpolarized before their decay. 

It is clear that hyperons 
in groups (a) and (b) can be polarized. 
Those in group (a) are polarized \cite{GH93,BL98}  
since the polarization of the initial quark $q_f^0$
can be transferred to 
such hyperon $H_i$'s in the fragmentation process.
We denote the probability for 
this polarization transfer by $t^F_{H_i,f}$ and 
call it polarization transfer factor 
from quark $q_f$ to hyperon $H_i$, 
where the superscript $F$ stands for fragmentation. 
Hyperons in group (b) can be polarized 
since such hyperons can inherit part of the polarization
of the parent hyperons in the decay process.
We denote the probability for 
this polarization transfer by $t^D_{H_i,H_j}$ and 
call it polarization transfer factor 
from hyperon $H_j$ to hyperon $H_i$, 
where the superscript $D$ stands for decay. 
In contrast, hyperons in groups (c) and (d) are unpolarized. 
Hence, we obtain that 
the polarization of the final hyperon is,
\begin{equation}
P_{H_i}={ {\sum\limits_f t^F_{H_i,f} P_f \langle n^a_{H_i,f}\rangle
+\sum\limits_{j} t^D_{H_i, H_j} P_{H_j} \langle n^b_{H_i, H_j}\rangle}
 \over
{\langle n^a_{H_i}\rangle +\langle n^b_{H_i}\rangle + 
\langle n^c_{H_i}\rangle +\langle n^d_{H_i}\rangle} }. 
\end{equation}
Here $P_f$ is the polarization of the initial quark $q_f^0$;
$\langle n^a_{H_i,f}\rangle$ is the average number of 
the hyperons which are directly produced and contain 
the initial quark of flavor $f$;
$P_{H_j}$ is the polarization of the hyperon $H_j$   
before its decay;
$\langle n^b_{H_i,H_j}\rangle$ is average numbers of $H_i$ hyperons
coming from the decay of $H_j$ hyperons 
which are polarized;
$\langle n^a_{H_i}\rangle(\equiv \sum\limits_f \langle n^a_{H_i,f}\rangle$),
$\langle n^b_{H_i}\rangle(\equiv \sum\limits_j \langle n^b_{H_i,H_j}\rangle)$,
$\langle n^c_{H_i}\rangle$ and $\langle n^d_{H_i}\rangle$
are average numbers of hyperons in group (a), (b), (c) 
and (d) respectively. 
These different factors can be calculated in the following way.

Longitudinal polarization of the 
initial quark $q_f^0$ is determined by 
the standard model for electroweak interactions.  
The polarization comes from the weak interactions 
and the results for $e^+e^-\to \gamma^*/Z\to q_f^0\bar q_f^0$ 
can be found e.g. in [\ref{AR80}], i.e.,
\begin{equation} 
P_f=-\frac{A_f(1+\cos^2\theta)+B_f\cos\theta}
          {C_f(1+\cos^2\theta)+D_f\cos\theta},
\end{equation}
where $\theta$ is the angle between the outgoing quark and 
the incoming electron, the subscript $f$ denotes the flavor 
of the quark, and
\begin{equation}
A_f=2a_fb_f(a^2+b^2)-2(1-{m_Z^2 \over s})Q_fab_f,
\end{equation}
\begin{equation}
B_f=4ab(a^2_f+b^2_f)-2(1-{m_Z^2 \over s})Q_fa_fb,
\end{equation}
\begin{equation}
C_f=\frac{(s-m_Z^2)^2+m_Z^2\Gamma^2_Z}{s^2}Q^2_f+
(a^2+b^2)(a^2_f+b^2_f)-2(1-{m_Z^2 \over s})Q_faa_f,
\end{equation}
\begin{equation}
D_f=8aba_fb_f-4(1-{m_Z^2 \over s})Q_fbb_f,
\end{equation}
where $m_Z$ and $\Gamma_Z$ are the mass and decay width of $Z$; 
\begin{equation}
a=\frac{-1+4\sin ^2\theta_W}{2\sin 2\theta_W},
\end{equation}
\begin{equation}
b=-\frac{1}{2\sin 2\theta_W},
\end{equation}
\begin{equation}
a_f=
 \left\{ \begin{array}{r@{\quad\quad}l} 
\frac{1-8\sin ^2\theta_W/3}{2\sin 2\theta_W}, & 
 \mbox {\rm for } f=u,c,t,\\ 
\frac{-1+4\sin ^2\theta_W/3}{2\sin 2\theta_W}, &
 \mbox {\rm for } f=d,s,b, 
\end{array} \right.
\end{equation}
\begin{equation}
b_f= 
 \left\{ \begin{array}{r@{\quad\quad}l} 
\frac{1}{2\sin 2\theta_W}, & 
 \mbox {\rm for } f=u,c,t,\\ 
-\frac{1}{2\sin 2\theta_W}, &
 \mbox {\rm for } f=d,s,b, 
\end{array} \right.
\end{equation}
are the vector and axial vector coupling constants 
of electron and quark to $Z$ boson, and 
$\theta_W$ is the Weinberg angle. 
To see the size and the $\theta$-dependence of $P_f$, 
we show in Fig.1 the numerical results of $-P_f$ 
as a function of $\cos\theta$ at LEP I energy.
We see clearly that all the quarks 
produced at the $e^+e^-$ annihilation vertices 
are longitudinally with significantly high polarizations. 
We see also that the magnitude of the polarization 
of the down-type quarks ($d, s$ and $b$) is large and 
varies little with the angle $\theta$,
while that of the 
up-type quarks ($u$ and $c$) is a little bit smaller 
and has a relatively larger variation 
in the whole range of $\cos\theta$. 
The latter increases with $\cos\theta$ from 
0.58 at $\cos\theta=-1$ to 
0.74 at $\cos\theta=1$. 
Averaging over $\theta$, we obtain, 
\begin{equation}
\langle P_f \rangle=\frac{\int P_f \sigma^0d\cos\theta}
{\int\sigma^0d\cos\theta}=-{{A_f}\over{C_f}},
\end{equation}   
where $\sigma^0=C_f(1+\cos^2\theta)+D_f\cos\theta$
is the angular variation of the unpolarized cross section.
In Fig.2, we show 
$-\langle P_f\rangle $ as a function 
of the total $e^+e^-$ center-of-mass energy $\sqrt{s}$.
We see that, 
at the LEP I energy, i.e. $\sqrt{s}=91GeV$,  
the polarizations of quarks
have the maximum negative value,
i.e., $\langle P_f\rangle=-0.67$ 
for $f=u,c$,
and  $\langle P_f\rangle=-0.94$
for $f=d,s,$ and $b$.
From the LEP I energy to the LEP II energy, 
the polarizations decrease a bit 
but it is still very large at the LEP II energy 
($\sqrt{s}=200GeV$), where we have
$\langle P_f\rangle=-0.26$ for $f=u,c$,
and  $\langle P_f\rangle=-0.8$
for $f=d,s,$ and $b$.

At LEP II energy, the $W^+W^-$ events, 
i.e. $e^+e^-\to W^+W^-\to q_{f1}\bar q_{f2}q_{f3}\bar q_{f4}$, 
are significant. 
It takes about 10\% of the whole events. 
The polarizations for the initial quarks 
in such events are different from those in 
$e^+e^-\to \gamma^*/Z^0\to q_{f }\bar q_{f }$, 
and should be considered separately in the calculations.
According to the standard model for electroweak interactions, 
the polarization of initial quark and that of initial anti-quark 
created in $W^+$ or $W^-$ decay vertex
are equal to -1.0 and 1.0, respectively.
There exists also another type of events, i.e. 
$e^+e^-\to Z^0Z^0\to q_{f1}\bar q_{f1}q_{f2}\bar q_{f2}$, 
which contribute at LEP II energy. 
But this contribution is very small (less than 1\%). 
We will not consider such events in the following calculations.

The fragmentation polarization 
transfer factor $t^F_{H_i,f}$ 
from the initial quark $q_f$ to hyperon $H_i$ 
is equal to the fraction of 
spin carried by the $f$-flavor-quark 
divided by the average number of quark of flavor $f$ 
in the hyperon $H_i$\cite{GH93,BL98}.
This fractional contribution to the  
hyperon spin from $f$-flavor-quark 
is different in the above-mentioned 
SU(6) or the DIS picture. 
The results in the SU(6) picture can easily be obtained 
from the wave functions. 
In the DIS picture, the fractional contribution 
of quarks of different flavors 
to the spin of a baryon in the $J^P={1\over2}^+$ octet
is extracted from $\Gamma_1^p\equiv \int^1_0 g_1^p(x) dx$
obtained in deep-inelastic
lepton-proton scattering experiments\cite{SPIN97}
and the constants $F$ and $D$ obtained
from hyperon decay experiments. 
The way of doing this extraction 
is now in fact quite standard 
(see, for example, the Appendix in [\ref{BL98}]).
The results for $\Lambda$, $\Sigma^0$ and $\Xi$ 
hyperons have been obtained in [\ref{Jaffe96}] and [\ref{BL98}]. 
Using exactly the same way, we obtain 
also the results for $\Sigma^\pm$. 
For completeness, we list the results for all the 
$J^P={1\over 2}^+$ hyperons in Table 1.

In the calculations, we take also the contributions from the 
decay of $J^P={3\over 2}^+$, i.e. the decuplet hyperons into account. 
The polarizations of such hyperons in the SU(6) picture 
can easily be calculated using the SU(6) symmetric wave functions. 
But, it is presently impossible to calculate them in the DIS picture 
since no DIS data is available for any one of the decuplet baryons.
Therefore, in the calculation 
we take them into account in the same way 
as those in the SU(6) picture.
 
The decay polarization transfer factor $t^D_{H_i,H_j}$ 
is determined by the decay process and is independent 
of the process in which $H_j$ is produced. 
The results for different decay processes are different 
and will be given for each individual hyperon 
in the following sections of this paper. 

After we obtain the results for $P_f$, $t^F_{H_i,f}$ 
and $t^D_{H_i,H_j}$, we can calculate $P_{H_i}$ 
if we know the average numbers $\langle n^a_{H_i,f}\rangle$, 
$\langle n^b_{H_i,H_j}\rangle$, $\langle n^c_{H_i}\rangle$, 
and $\langle n^d_{H_i}\rangle$ for hyperons $H_i$ 
from the different origins. 
These average numbers are determined 
by the hadronization mechanism and should be 
independent of the polarization of the initial quarks.
Hence, we can calculate them using the a hadronization 
model which give a good description of the unpolarized 
data for multiparticle production in high energy reactions. 
Presently, such calculations can only be carried out using 
a Monte-Carlo event generator. 
We use the Lund string fragmentation model\cite{AGIS83}
implemented by JETSET \cite{Sjo86} in the following.

\subsection{$\Lambda$ polarization in the average events}

Among all the $J^P={1\over 2}^+$ hyperons, 
$\Lambda$ is most copiously produced. 
Furthermore, 
the spin structure of $\Lambda$ in the $SU(6)$ picture
is very special, which makes it play a very special role 
in distinguishing the SU(6) and the DIS pictures. 
In the $SU(6)$ picture, 
spin of $\Lambda$ is completely carried by the $s$ valence quark,
while the $u$ and $d$ quarks have no contribution.
Since the initial $s$ quark produced 
in the annihilation of the initial $e^+e^-$ 
takes the maximum negative polarization,  
$|P_\Lambda|$ obtained using 
the SU(6) picture is the maximum among 
all the different models. 
In contrast, in the DIS picture, 
the $s$ quark carries only about $60\%$ of the $\Lambda$
spin, while the $u$ or $d$ quark each carries about $-20\%$ 
(see Table 1).
The resulting $|P_\Lambda|$ should be substantially 
smaller than that obtained in the $SU(6)$ picture.
Comparing the maximum with experimental results 
provide us a good test of the validity of the picture. 

$\Lambda$ is also the lightest hyperon, 
so the final $\Lambda$'s produced 
in high energy reactions 
contain contributions from 
the decays of many different heavier hyperons.  
There are three octet,
i.e. $\Sigma^0$, $\Xi^0$ and $\Xi^-$,
and five decuplet,
i.e., $\Sigma(1385)^{\pm,0}$ and $\Xi(1530)^{-,0}$, 
hyperons that can decay to $\Lambda$.
The polarization transfer in 
$\Sigma^0\to \Lambda \gamma$ 
has been studied in [\ref{Gatto58}], the 
result for the polarization transfer 
factor $t^D_{\Lambda,\Sigma^0}$ is $-1/3$.
$\Xi \to \Lambda \pi$ is a parity
non-conserving decay and is dominated by S-wave. 
The polarization transfer 
$t^D_{\Lambda,\Xi}$ 
for this process is equal to $(1+2\gamma)/3$,
where $\gamma = 0.87$ can be found 
in Review of Particle Properties [\ref{C98}]. 
The decay process from the decuplet hyperon 
to octet hyperon and a pseudo-scalar meson $\pi$ 
such as $\Sigma(1385) \to \Lambda \pi$ or
$\Xi(1530) \to \Xi \pi$ is 
dominated by the $P$ wave\cite{GH93}, 
and the octet hyperon will get the same polarization as 
that of the initial decuplet hyperon,  
i.e., $t^D=1$.
For explicity, we list all these 
results for $t^D_{\Lambda,H_j}$ in table.2.
There are also some contributions from
the decays of open charm or open beauty baryons. 
The polarization transfer factors in these decay processes 
are unfortunately unknown yet. 
This is a theoretical uncertainty in the calculations.
However, according to the materials in 
Review of Particle Properties\cite{C98}, 
we do know that each of these open charm or open beauty baryons 
can decay to $\Lambda$ through many different channels.  
The contributions to $\Lambda$ polarization
in these different channels may be quite different.   
We expect that the net contribution 
from all these channels cannot be large. 
We will just neglect it in our calculations.

Using JETSET and PYTHIA,
we obtain the average numbers $\langle n^a_{\Lambda,f}\rangle$,
$\langle n^b_{\Lambda,H_j}\rangle$, $\langle n^c_{\Lambda}\rangle$,
and $\langle n^d_{\Lambda}\rangle$ at $\sqrt{s}=91$ GeV and those 
at $\sqrt{s}=$200 GeV,respectively.
We show them in Fig.3 and Fig.4.
The results at $\sqrt{s}=91$ GeV are of course the same as those 
in [\ref{GH93}] and [\ref{BL98}]. 
From these figures, we see clearly that 
contribution to $\Lambda$ from the events originating  
from the initial $s$ quarks play the most important role, 
in particular for large $z$.
For example, for $z>0.4$,   
it gives about 70\% of the whole $\Lambda$'s,  
while those from the events initiated by 
$u$ or $d$ take only 10\% respectively. 

Using the results for the $P_f$'s, 
the $t^D_{H_i,f}$'s, and the $t^D_{H_i,H_j}$'s 
mentioned above, we obtain 
the longitudinal polarization of $\Lambda$ 
as shown in Fig.5.
A comparison of those results at LEP I 
with the ALEPH data \cite{ALEPH96} 
and the OPAL data \cite{OPAL98}
shows that the data\cite{ALEPH96,OPAL98} 
of both groups agree better with 
the calculated results based on the $SU(6)$ picture. 
But, these available data\cite{ALEPH96,OPAL98}  
are still far from accurate and enormous enough 
to make a decisive conclusion. 
Further complementary measurements are needed.

At the LEP II energy, $W^+W^-$ events take 
about 10\% of the whole events.  
We expect that such events give a even more 
significant contribution
to $\Lambda$ polarization,
since the polarization of the initial quark 
at $W^+$ or $W^-$ decay vertex is 100\%, 
which is larger than those from the 
$e^+e^-\to \gamma^*/Z^0\to q_f^0\bar q_f^0$.
Since there are two initial quarks and 
two initial anti-quarks in $W^+W^-$ events, 
the energy of each of them should be much smaller 
than those in the $e^+e^-\to \gamma^*/Z^0\to q_f^0\bar q_f^0$ events. 
Hence, such events should contribute mainly 
in relatively smaller $z$ regions. 
Adding all the different together, we obtain \cite{ft}
$P_\Lambda$ at LEP II energy shown in Fig.5.
We see that, compared with those obtained 
at the LEP I energy, 
$|P_\Lambda|$ at LEP II energy 
is a little larger at small $z$ region,
and is a little smaller at large $z$ region.
This is consistent with the above-mentioned 
qualitative expectations and can be checked experimentally. 

\subsection{$\Lambda$ polarization in different subsamples of events}

In the two models we discussed above, 
$\Lambda$ polarization comes solely from the hyperons 
which contain the initial quark created at the 
$e^+e^-$ annihilation vertex. 
The contributions are very much different for 
the hyperons which contain the initial $s$ 
from those which contain the initial $u$ or $d$ quarks. 
There is also a big differences 
between the contribution from the hyperons 
which contain the initial quarks and that from 
those which do not. 
It is thus clear that we can get a further 
check of the two pictures  
if we study the $\Lambda$ polarization in 
events which originate from
the initial $u$ or $d$ quark or those 
which originate from the initial $s$ quark separately.
We should also get a very sensitive check to different 
pictures if we study only those $\Lambda$'s which 
contain the initial quarks.

In this connection, it should also be mentioned 
that there exist another different 
type of models\cite{Kotz98,Ma2000} 
for spin transfer in fragmentation processes. 
In contrast to that described in subsection \ref{subs:method},  
in these models, no distinction is made between 
the hyperons which contain the initial quarks 
or those which do not contain the initial quarks.  
Some of them\cite{Ma2000} even do not distinct 
those which are directly produced or 
those which come from heavier hyperon decays. 
In other words, in these models, 
there is no distinction between hyperons from 
the groups (a), (b), (c) and (d).   
It is simply assumed that a 
``reciprocity relation'' is valid between 
the fragmentation function for the final hyperons 
and the corresponding quark distribution functions 
in the hyperons. 
They are taken as proportional to each other, 
with a proportional constant which is common for  
quarks of different flavors.
The hyperon polarization can easily be calculated 
in such models if the spin-dependent quark 
distribution functions are known. 
The obtained results depend obviously 
very much on the spin-dependent quark distributions, 
which are used as theoretical input. 
Since the spin dependent quark distributions 
in hyperons are still poorly known yet, 
the obtained results can be very much different 
from each other if different sets of 
quark distributions are used.
But, since the (polarized) fragmentation function 
from a given quark $q^0_f$ to a given hyperon $H_i$ 
depends only on the fractional momentum $z$ of the 
quark carried by the produced hyperon,   
the obtained polarization at a given $z$ 
should be completely the same if we study 
only those hyperons which contain 
the initial quarks or all of them 
in events originating from a given type of initial quark. 

It is unfortunately impossible to 
make a complete separation 
of hyperons in these different groups in experiments. 
However, it should be possible to separate the events 
into different subsamples, in each of 
which the hyperons from one group dominate. 
Measuring $\Lambda$ polarization in 
such subsamples of events should give further complementary 
checks of the different models mentioned above.
Using the Monte-Carlo event generator JETSET \cite{Sjo86}, 
we can study the various possibilities in this direction. 
We note that, if a $\Lambda$ originates from the initial $s$ quark, 
the leading particle in the opposite direction should 
contain the $\bar s$ produced 
in the $e^+e^-$ annihilation vertex.   
Hence, we choose an event sample according to the following 
criteria: 

(i) $\Lambda$ is the leading in one direction; 

(ii) the leading particle in the opposite direction is $K^+$. 

\noindent
We expect that such $\Lambda$'s should mainly have the 
origin (a) mentioned in the subsection \ref{subs:method}.
In Fig.6, we show the results obtained from JETSET 
for the average numbers of $\Lambda$'s of different origins. 
We see that such leading $\Lambda$'s indeed 
mainly originate from the initial $s$ quark. 
The contribution from the initial $u$ or $d$ quark
is indeed substantially small.  
It takes only about 3\% of such $\Lambda$'s. 
In particular, in $z>0.5$ region there is almost no
contribution from the initial $u$ or $d$ quark at all.

Using Eq.(1), we calculated $P_\Lambda$ for 
the events under the conditions (i) and (ii) 
using the SU(6) and the DIS pictures.
The comparison of the obtained results 
with those obtained in the average events 
are shown in Fig.7.
We see that there is indeed a significant difference 
between the results obtained for such particularly 
chosen events and those for the average events  
in particular in the region of $z>0.3$. 
Checking such differences by measuring 
$P_\Lambda$ for such events can be helpful in 
distinguishing the validity of different pictures 
for the spin transfer in fragmentation processes. 

\section{Longitudinal polarization of other $J^P={1\over2}^+$ hyperons}

The production rates for other octet hyperons 
are smaller than that for $\Lambda$ so 
the statistic errors should be larger for 
the polarizations of these hyperons. 
On the other hand, decay contributions 
from heavier hyperons to these hyperons are also 
much less significant than that in case of $\Lambda$. 
Hence, the contaminations from the decay processes 
are much smaller. 
These conclusions can easily be checked using 
a Monte-Carlo event generator for $e^+e^-$ 
annihilation into hadrons.
In Fig.8, we show the results obtained from JETSET
for these hyperons compared with those for $\Lambda$. 
We see that, the production rate for $\Sigma^+$ or $\Xi^0$ 
is less than $20\%$ of that for $\Lambda$. 
On the other hand, 
we see also that the contribution 
from heavier hyperon decays is also much smaller.  
For example, for $\Sigma^+$'s, the decay contribution 
takes only about $7\%$ of the total rate. 
The situations for $\Sigma^-$, $\Xi^0$, and $\Xi^-$
are similar to that for $\Sigma^+$. 
Most of them are directly produced.
Hence, the theoretical uncertainties in the 
calculations for these hyperons are much smaller.
The study of polarizations of these hyperons 
should provide us with good complementary tests 
of different pictures.
In this section, we calculate 
the longitudinal polarizations of these hyperons,
i.e., $\Sigma^+$, $\Sigma^-$, $\Xi^0$ and $\Xi^-$,
in $e^+e^-$ annihilation at LEP I and LEP II energies.

The calculations are quite similar to those 
for $\Lambda$ production. 
The general procedure is the same and 
has been outlined in last section. 
The polarization transfer factor $t^F_{H_i,f}$ 
are given in table 1.  
The situation for decay contributions is quite simple:  
There are only contribution from the corresponding 
decuplet hyperon decays. 
More precisely, $\Sigma$ contains 
only decay contribution from $\Sigma(1385)$ 
and $\Xi$ has the decay contribution from $\Xi(1530)$. 
Both decay processes are strong decays and 
are dominated by the P-wave. 
The polarization transfer is the same as that 
in $\Sigma(1385)\to \Lambda \pi$, i.e.  
$t^D_{\Sigma,\Sigma(1385)}=t^D_{\Xi,\Xi(1530)}=1$.  

Before we present the numerical results of the 
calculations, we would like to note the 
following qualitative expectations. 

First, although the spin structure 
of $\Sigma^+$ and that of $\Sigma^-$ or 
that of $\Xi^0$ and that of $\Xi^-$ 
are symmetric under the exchange of $u$ and $d$ 
(c.f. Table 1), 
their polarizations should be quite different from each other. 
This is because 
the polarization of the initial $u$ 
and that of the initial $d$ 
produced at the $e^+e^-$ annihilation vertices 
are quite different.  
This can be seen clearly in Fig.1, where 
we see that the magnitude of the polarization of $u$ quark
is smaller than that of $d$ quark. 

Second, from Table 1,
we see that the contributions from the two different 
flavors in these hyperons ($\Sigma^{+,-}$, $\Xi^{0,-}$)
are quite different: they have different signs 
and  different magnitudes,  
and the differences are
also different in the SU(6) or the DIS picture. 
This implies that their contributions to 
hyperon polarizations are also quite different. 
They should have different signs and different magnitudes.
These differences make the situation very interesting. 
We should have different expectations for 
different hyperons.   
Even the signs of polarizations of some hyperons
in the two different pictures can be different from each other.  

Using JETSET and PYTHIA, we calculated the different contributions
to $\Sigma^+$, $\Sigma^-$, $\Xi^0$ and $\Xi^-$ 
from all the different sources discussed above
at LEP I and LEP II, respectively.
The results at LEP I energy are shown in Fig.9. 
Those at LEP II energy are similar and are not shown. 
From this Fig.9, we see that the $\Sigma^+$'s
mainly come from the initial $s$ and the initial $u$-quark, 
and that the $\Sigma^-$'s come mainly from
the initial $s$ and the initial $d$-quark, 
and that these two contributions are 
comparable with each other in particular for $\Sigma^-$'s.
In contrast, the source for $\Xi^{0,-}$ is very pure: 
They come predominately from the initial $s$-quarks. 
All the others are negligible.
  
Using Eq.(1) and the results shown in Fig.9,
we calculated the hyperon polarizations,
$P_{\Sigma^+}$, $P_{\Sigma^-}$, $P_{\Xi^0}$ and $P_{\Xi^-}$. 
The obtained results are shown in Fig.10.
From these results, we see clearly that 
the polarizations are different for these different hyperons, 
and that there are considerably large differences
between the results using the $SU(6)$ picture
and those using the DIS picture.
In particular,
we see that $P_{\Sigma^+}$ 
in the $SU(6)$ picture has opposite sign 
to that in the DIS picture.
But their magnitudes are relatively small.
The magnitude of $P_{\Xi^0}$ and that of $P_{\Xi^-}$ 
are considerably larger, in particular in the large $z$ region.
This is because the contribution to $\Xi^0$ or $\Xi^-$ 
is very clear. 
They come mainly from the initial $s$-quark. 
There is thus little theoretical uncertainty 
in calculating $P_{\Xi}$. 
Hence, although the statistics 
may be much worse than $\Lambda$, 
there are still good reasons to study $P_{\Xi}$ 
in $e^+e^-\to \Xi+X$ at LEP energies.

\vskip 1.0cm

We thank G\"osta Gustafson, Li Shi-yuan, Wang Qun, Xie Qu-bing 
and other members in the theoretical particle physics group of 
Shandong University for helpful discussions.  
This work was supported in part by the National Science Foundation 
of China (NSFC) and the Education Ministry of China.

\widetext

\begin{table}
\caption{Fractional contributions $\Delta U$, $\Delta D$ and $\Delta S$ 
of the light flavors to the spin of baryons in the $J^P={1\over 2}^+$ 
octet calculated using the $SU(6)$ picture and those obtained using 
the data for deep inelastic lepton-nucleon scattering and those for 
hyperon decay under the assumption that SU(3) flavor symmetry is valid.  
The first column shows the obtained expressions in terms of 
$\Sigma$, $F$ and $D$. 
The $SU(6)$ picture results are obtained by inserting 
$\Sigma=1, F=2/3$ and $D=1$ into these expressions 
and those in the third column are obtained by inserting 
$\Sigma =0.28$, obtained from deep inelastic lepton-nucleon 
scattering experiments [\ref{SPIN97}], 
and $F+D=g_A/g_V=1.2573, F/D=0.575$  
obtained [\ref{C98},\ref{CR93}] from the hyperon decay experiments. }
\begin{tabular}{l||c|c|c||c|c|c}
\hline 
 &\multicolumn{3}{c||}{$\Lambda$} &\multicolumn{3}{c}{$\Sigma^0$}\\ \hline 
& & $SU(6)$ & DIS & & $SU(6)$ & DIS  \\ \hline
$\Delta U$ & $\frac{1}{3} (\Sigma-D)$  & 0    & -0.17 & 
             $\frac{1}{3} (\Sigma+D)$  & 2/3  & 0.36  \\ \hline  
$\Delta D$ & $\frac{1}{3} (\Sigma-D)$  & 0    & -0.17 & 
             $\frac{1}{3} (\Sigma+D)$  & 2/3  & 0.36  \\ \hline
$\Delta S$ & $\frac{1}{3} (\Sigma+2D)$ & 1    & 0.62  & 
             $\frac{1}{3} (\Sigma-2D)$ & -1/3 & -0.44 \\ \hline \hline 
&\multicolumn{3}{c||}{$\Sigma^+$} &\multicolumn{3}{c}{$\Sigma^-$}\\ \hline
& & $SU(6)$ & DIS & & $SU(6)$ & DIS  \\ \hline
$\Delta U$ & $\frac{1}{3} (\Sigma+D)+F$  & 4/3    & 0.82 &
             $\frac{1}{3} (\Sigma+D)-F$  & 0  & -0.01  \\ \hline
$\Delta D$ & $\frac{1}{3} (\Sigma+D)-F$  & 0    & -0.10 &
             $\frac{1}{3} (\Sigma+D)+F$  & 4/3  & 0.82  \\ \hline
$\Delta S$ & $\frac{1}{3} (\Sigma-2D)$ & -1/3    & -0.44  &
             $\frac{1}{3} (\Sigma-2D)$ & -1/3 & -0.44 \\ \hline \hline
 &\multicolumn{3}{c||}{$\Xi^0$} &\multicolumn{3}{c}{$\Xi^-$}\\ \hline 
&  & $SU(6)$ & DIS & & $SU(6)$ & DIS  \\ \hline 
$\Delta U$ & $\frac{1}{3} (\Sigma-2D)$  & -1/3 & -0.44 & 
             $\frac{1}{3} (\Sigma+D)-F$ & 0    & -0.10 \\ \hline  
$\Delta D$ & $\frac{1}{3} (\Sigma+D)-F$ & 0    & -0.10 & 
             $\frac{1}{3} (\Sigma-2D)$  & -1/3 & -0.44 \\ \hline
$\Delta S$ & $\frac{1}{3} (\Sigma+D)+F$ & 4/3  &  0.82 & 
             $\frac{1}{3} (\Sigma+D)+F$ & 4/3  &  0.82 \\ \hline 
\end{tabular}
\end{table}

\newpage 
\begin{table}
\caption{The polarization transfer factors 
$t^D_{H_i,H_j}$ from hyperon $H_j$ to 
hyperon $H_i$ in the decay processes.}
\begin{tabular}{c||c} \hline 
 & $t^D_{H_i,H_j}$ \\ \hline 
$\Sigma^0 \to \Lambda \gamma$ & -1/3    \\ \hline  
$\Xi \to \Lambda \pi$  & $(1+2\gamma)/3$     \\ \hline  
$\Sigma(1385) \to \Lambda \pi$ & 1      \\ \hline
$\Xi(1530) \to \Xi \pi$  & 1     \\ \hline
$\Sigma(1385)\to \Sigma \pi$ & 1    \\ \hline  
\end{tabular}
\end{table}

\newpage 

\noindent
\vskip 0.6truecm
\noindent

\vskip 0.3cm
\noindent

\begin{figure}[ht]
\psfig{file=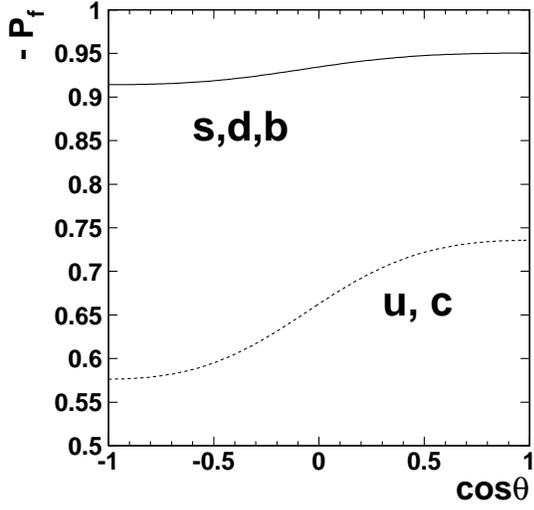,width=8cm}
\caption{Longitudinal polarization $-P_f$ 
of quark $q_f$ produced at the $e^+e^-$ annihilation vertex 
as a function of $\cos\theta$ at $\sqrt s=91GeV$.}
\label{fig1}
\end{figure}

\begin{figure}[ht]
\psfig{file=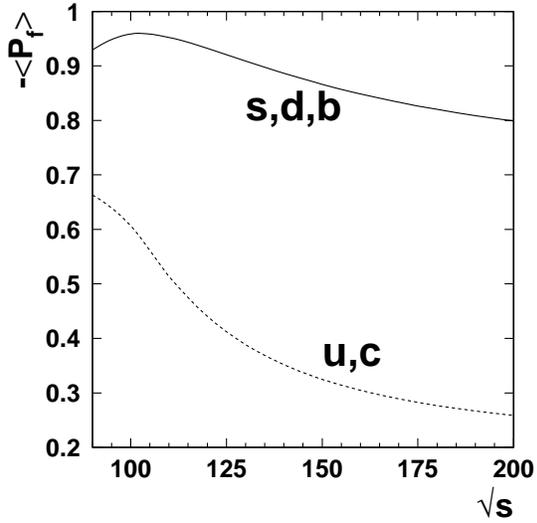,width=8cm}
\caption{Average values $-\langle P_f \rangle$ of the longitudinal 
polarization of quark $q_f$ produced at the $e^+e^-$ annihilation vertex 
as a function of the total c.m. energy $\sqrt s$ of the $e^+e^-$ system.}
\label{fig2}
\end{figure}

\begin{figure}[ht]
\psfig{file=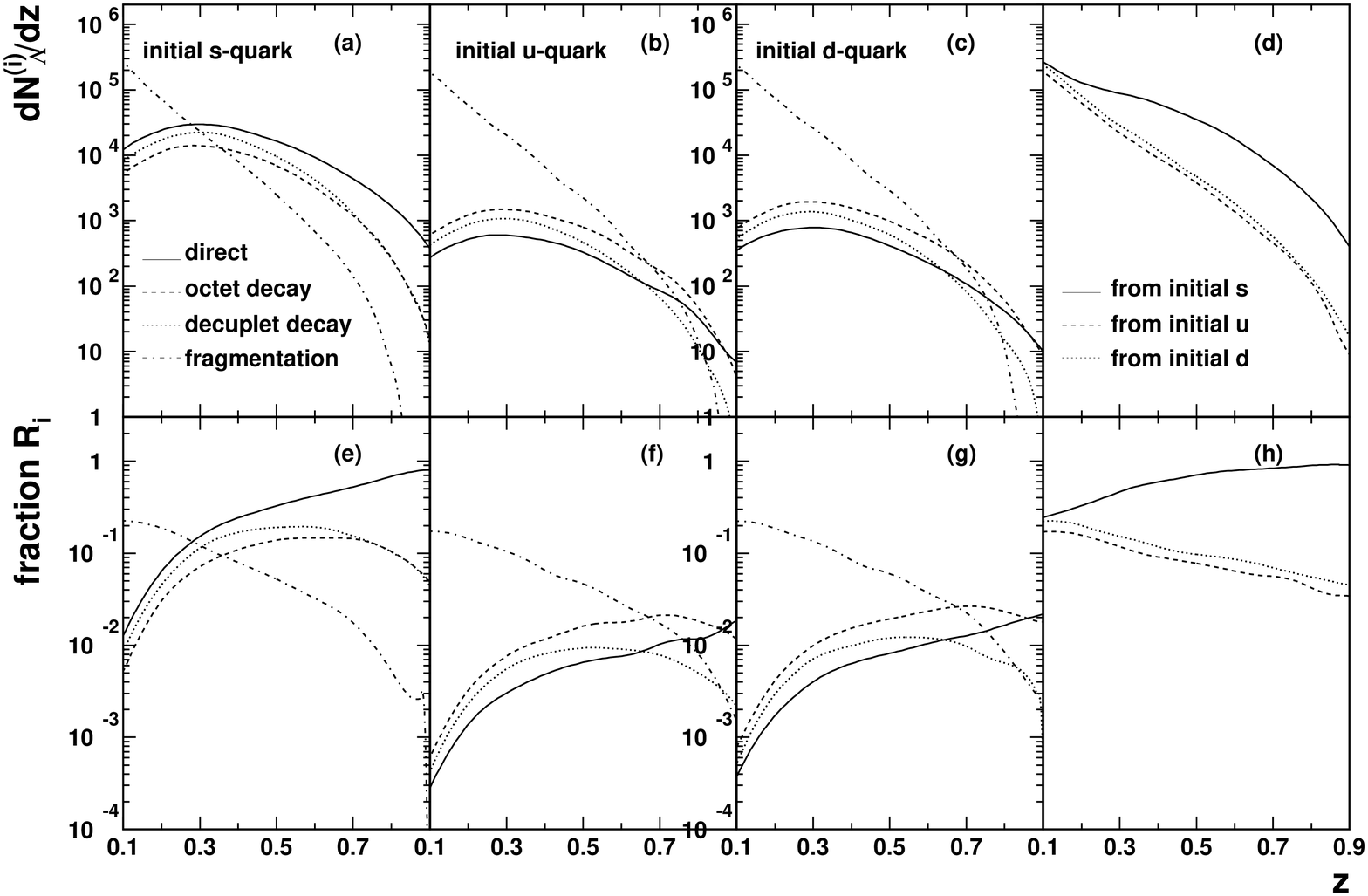,width=18cm}
\caption{Comparison of different contributions 
to $\Lambda$ in events originating
from the initial $s$, $u$, or $d$ quark
as functions of $z\equiv 2p/\sqrt{s}$ at LEP I energy. 
In Figs.(a), (b) and (c), we see the four types of 
contributions in initial $s$, $u$ or $d$ events respectively. 
Here, ``directly'' denotes those which 
are  directly produced and contain the initial quark; 
``octet decay'' and ``decuplet decay'' denote respectively 
those from the decay of the octet   
and the decuplet hyperons which contain the initial quark; 
``fragmentation'' denotes those directly produced 
but do not contain the initial $q^0$ plus those 
from unpolarized hyperon decay. 
In Figs. (e), (f) and (g), we see the fractions $R_i$ 
which are the ratios of 
the corresponding contributions to the sums of all these 
different contributions. 
In Figs. (d) and (h), we see the total contributions 
from events with initial $s$, $u$ or $d$ 
and the corresponding fractions respectively. 
}
\label{fig3}
\end{figure}

\begin{figure}[ht]
\psfig{file=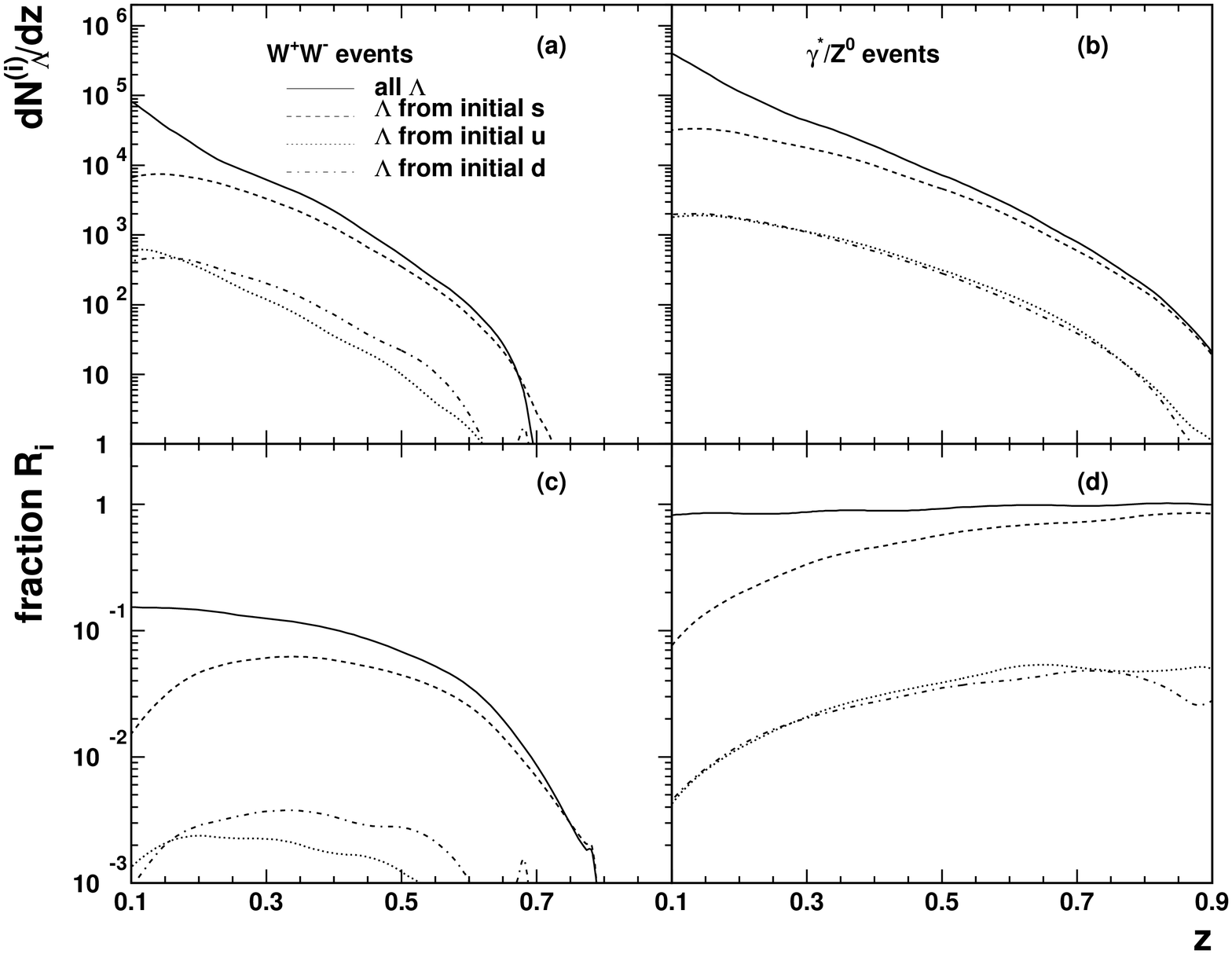,width=14cm}
\caption{Comparison of the different contributions to 
$\Lambda$ in $e^+e^-\to W^+W^-\to \Lambda+X$ events 
and those in $e^+e^-\to \gamma^*/Z^0\to \Lambda+X$ events 
at LEP II energy.
In Figs.(a) and (d), 
we see the contributions from initial 
$s$, $u$ or $d$ quark in these two classes of events. 
Here, the contributions include only 
those contributing to $\Lambda$ polarization, i.e.,  
they are the sums of those 
which are directly produced and contain the initial quarks 
with those from polarized heavier hyperon decays.
Figs.(c) and (d) represent 
the corresponding ratios 
to all $\Lambda$ in $e^+e^-\to \Lambda +X$. }
\label{fig4}
\end{figure}

\begin{figure}[ht]
\psfig{file=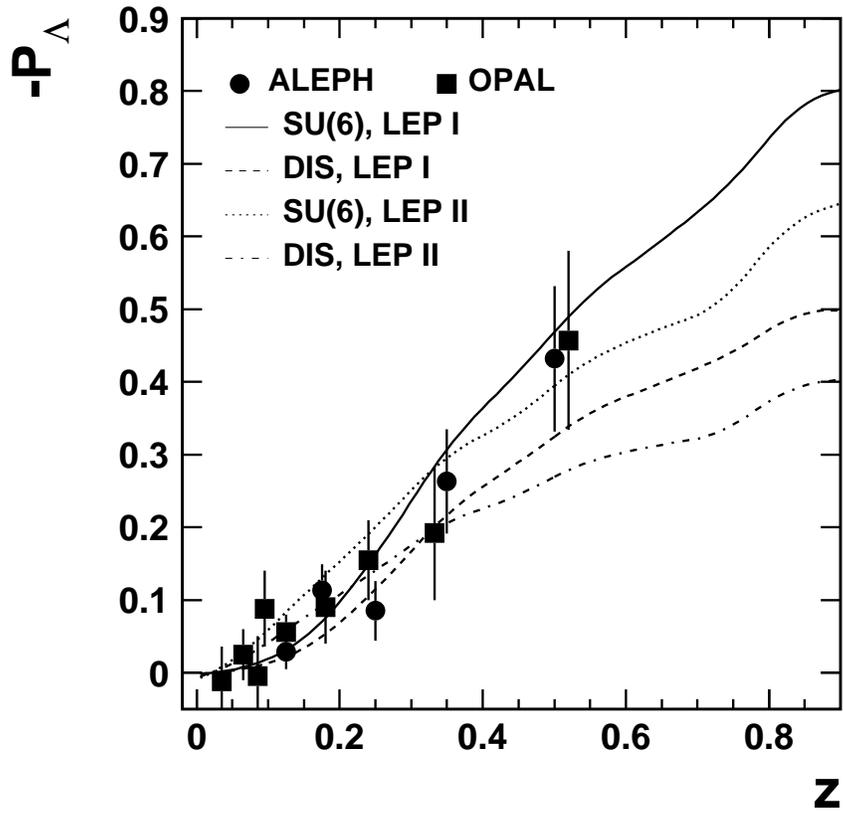,width=14cm}
\caption{
Longitudinal $\Lambda$ polarization, $P_\Lambda$,
in $e^+e^-\to \Lambda +X$ at LEP I and LEP II energies
as a function of $z$.
The data of ALEPH and those of OPAL are taken from
[\ref{ALEPH96}] and [\ref{OPAL98}] respectively.
}
\label{fig5}
\end{figure}

\begin{figure}[ht]
\psfig{file=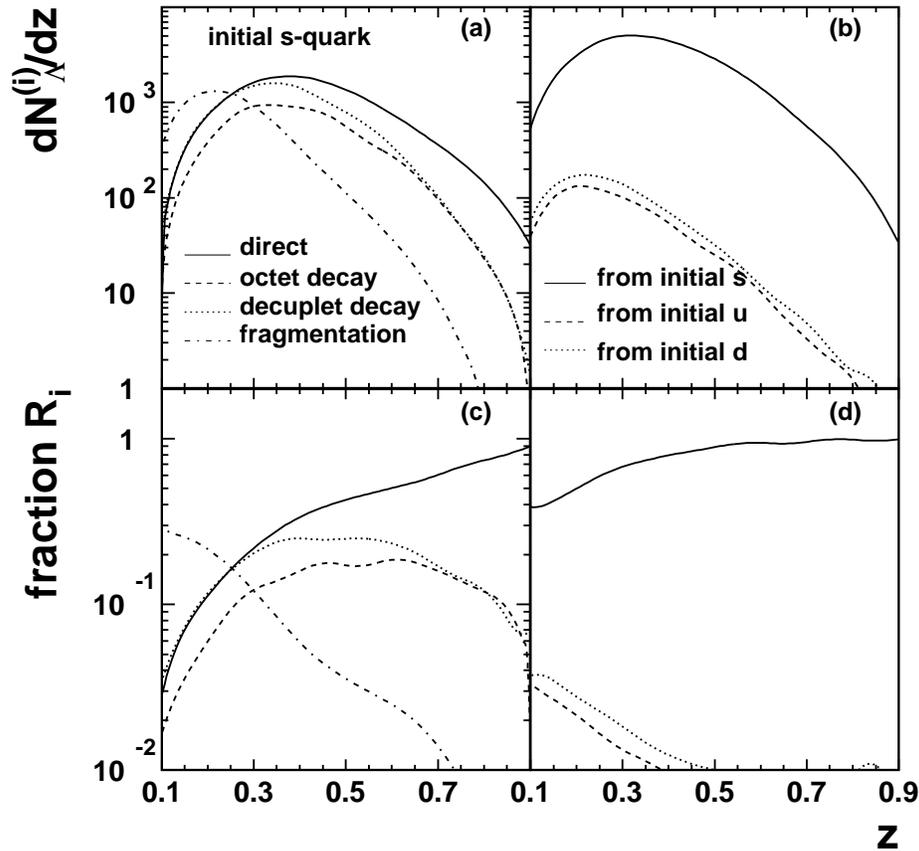,width=14cm}
\caption{
Different contributions to $\Lambda$ in the subsample of events 
where $\Lambda$ is the leading particle in one direction 
and the leading in the opposite direction is $K^+$ at LEP I energy.
In (a), we see the different contributions 
in the events with initial $s$-quarks. 
The means of the four curves are the same as those in Fig.4a. 
In (b), we see the contributions from 
the initial $s$, $u$ or $d$ events. 
In (c) and (d), we see the corresponding fractions 
of the contributions
in (a) and (b) to all $\Lambda$ in the subsample of events.
}
\label{fig6}
\end{figure}

\begin{figure}[ht]
\psfig{file=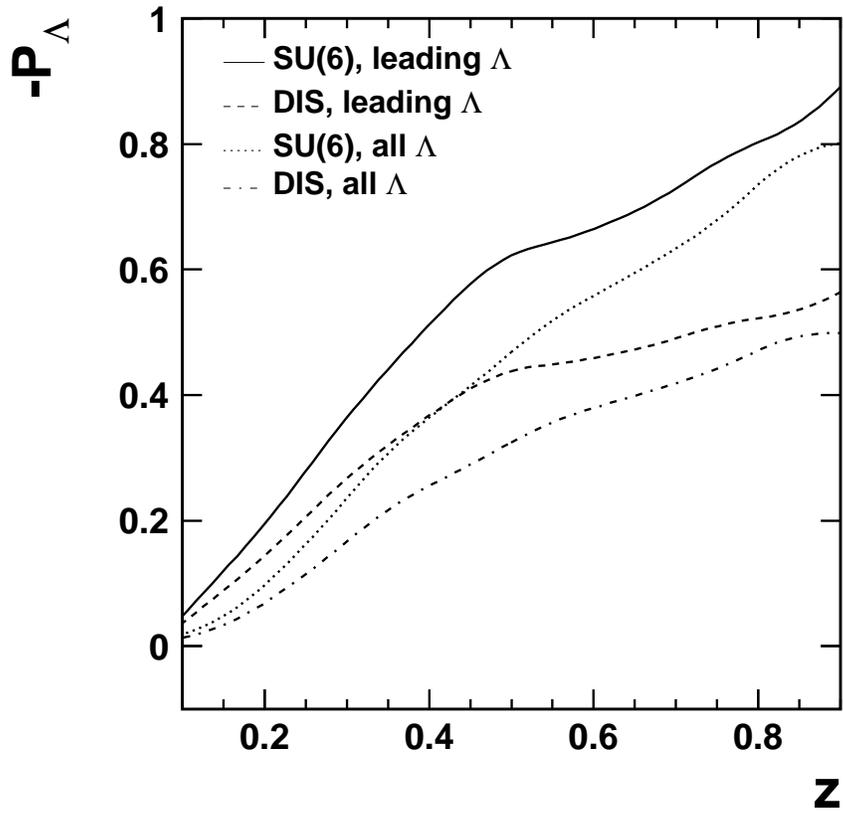,width=14cm}
\caption{ Comparison of $P_\Lambda$ as a function of $z$
in the subsample of events where $\Lambda$ 
is the leading in one direction
and $K^+$ is the leading in the opposite direction
with that in the average events at LEP I energy.
}
\label{fig7}
\end{figure}

\begin{figure}[ht]
\psfig{file=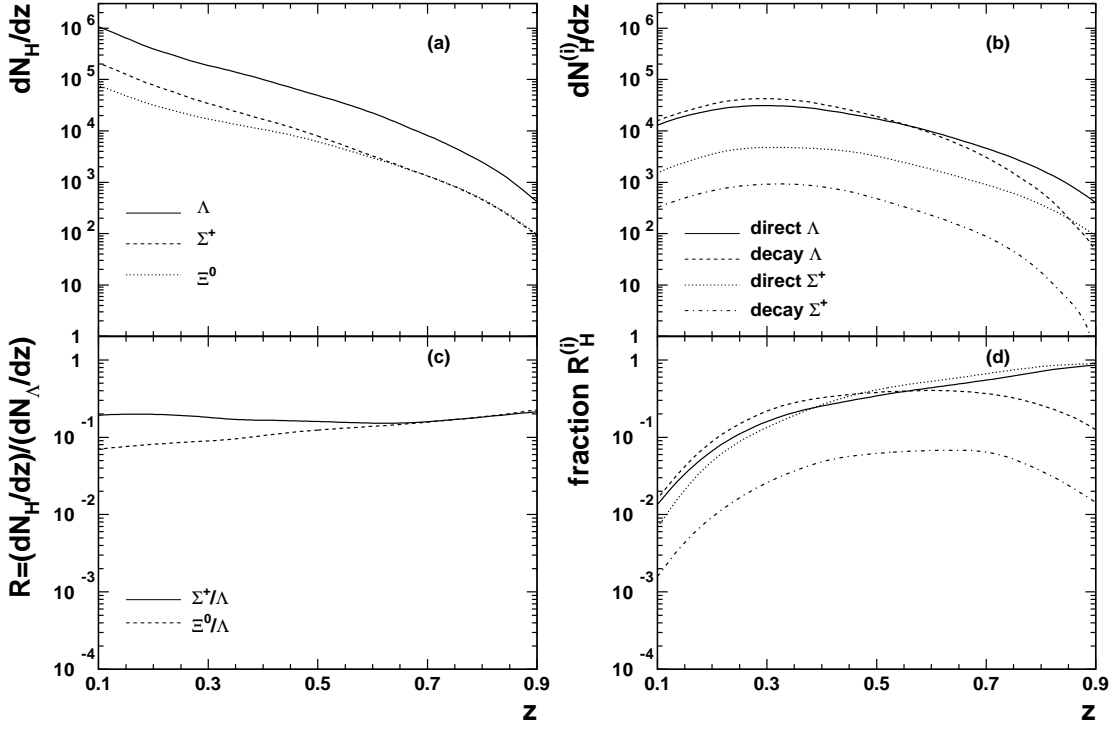,width=16cm}
\caption{ 
Comparison of the different contributions 
to $\Lambda$, $\Sigma^+$ or $\Xi^0$ 
in $e^+e^-$ annihilation at LEP I energy.
In (a) and (c), we see a comparison of the production rates of 
$\Sigma^+$, $\Xi^0$ to that of $\Lambda$. 
In (b), we see a comparison of the two different classes of 
contributions to $\Lambda$ and those to $\Sigma^+$. 
Here, ``directly'' represents the contribution 
from those which are directly produced and contain the initial quarks;
``decay'' denotes the contribution from 
polarized heavier hyperon decay. 
In (d), we see the ratios of contributions shown in (b) 
to all $\Lambda$ or all $\Sigma^+$ respectively.
}
\label{fig8}
\end{figure}

\begin{figure}[ht]
\psfig{file=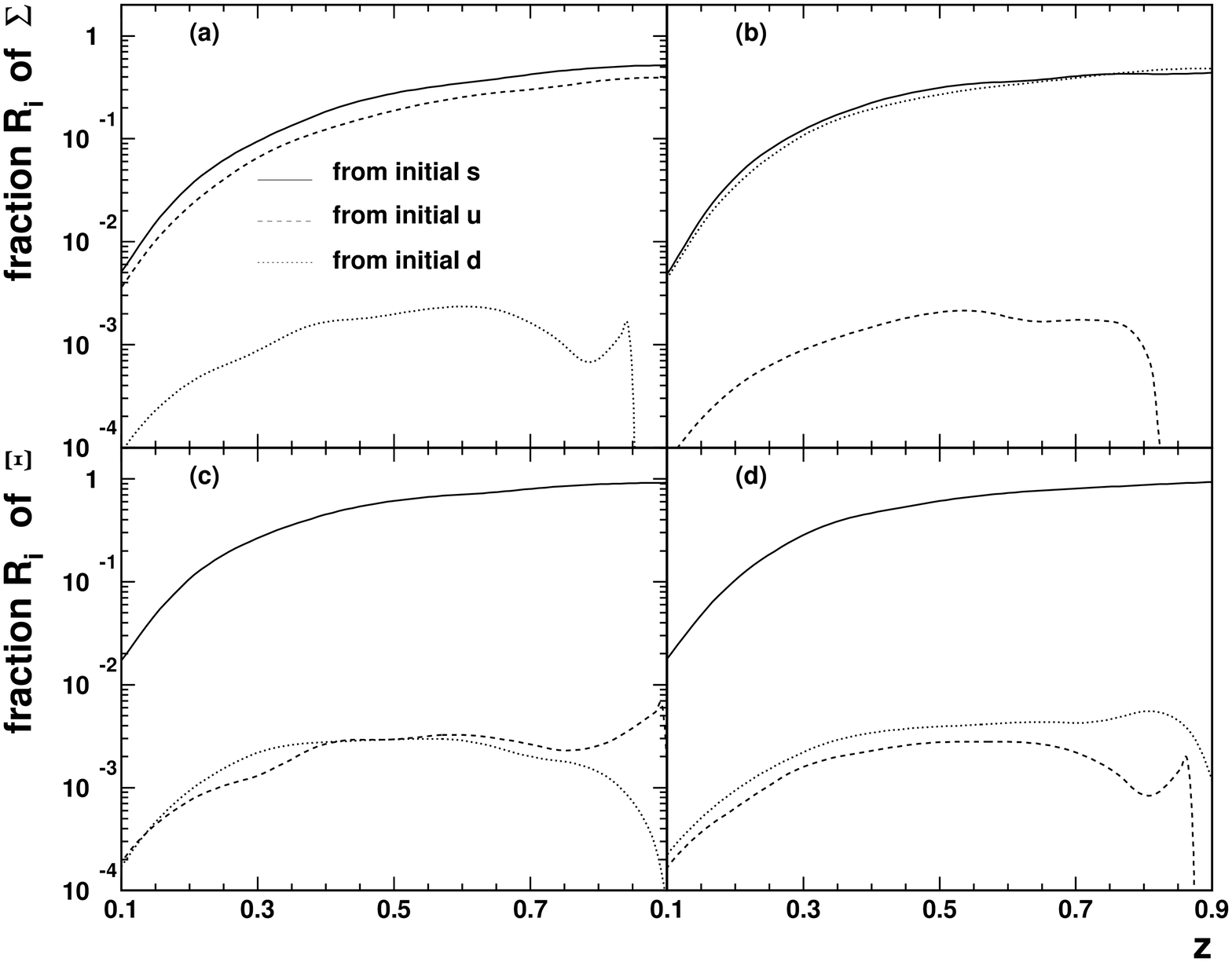,width=16cm}
\caption{ 
Contributions to $\Sigma^+$, $\Sigma^-$, $\Xi^0$ and $\Xi^-$ 
from the different sources in $e^+e^-$ annihilation at LEP I energy. 
Here, only those which contribute to hyperon polarization are included,  
i.e., the curves represent the corresponding sums of those    
which are directly produced and contain the initial $q_i^0$ 
with those from the decays of the polarized heavier hyperons.
}
\label{fig9}
\end{figure}

\begin{figure}[ht]
\psfig{file=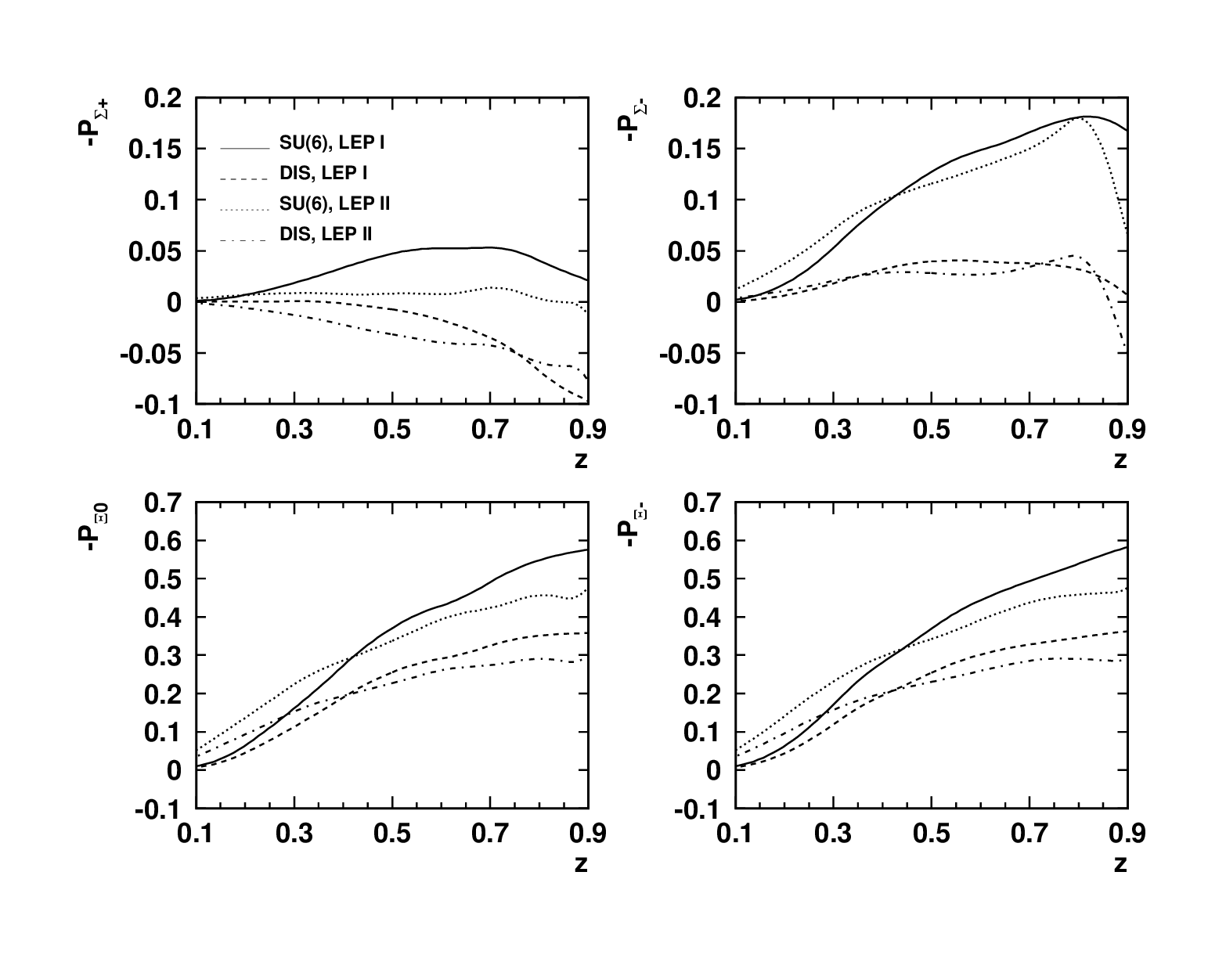,width=18cm}
\caption{Calculated results of 
the longitudinal polarizations of
$\Sigma^+$, $\Sigma^-$, $\Xi^0$ and $\Xi^-$
in $e^+e^-$ annihilation at LEP I and LEP II energies
as functions of $z$ from the two different pictures.}
\label{fig10}
\end{figure}


\begin {thebibliography}{99}
\bibitem{ft} It should be mentioned that, 
at the LEP II energy, 
the magnitude of the polarization of 
the initial quark produced 
in $e^+e^-\to \gamma^*/Z^0\to q_{f }\bar q_{f }$
can be influenced by the initial state radiation. 
Such initial state radiation reduces the effective 
energy of the $e^+e^-$ at the annihilation vertex 
thus the polarization of $P_f$ in such events should take 
the value at the corresponding lower energy. 
But, since $P_f$ varies only slowly with the energy, 
this effect cannot be large. 
We neglect it in the calculations.
\end{thebibliography}
\end{document}